\documentclass[floatfix,aps,twocolumn,showpacs,nofootinbib,superscriptaddress]{revtex4}

\usepackage{graphicx}
\usepackage{epsfig}

\usepackage{amsmath}
\usepackage{amsfonts}
\usepackage{amssymb}
\usepackage{bm}
\usepackage{mathtools}

\usepackage{hyperref}
\usepackage{dsfont}
\usepackage{lipsum}

\usepackage{tikz}
\usepackage{circuitikz}
\usetikzlibrary{arrows, shapes}
\usetikzlibrary{external}
\tikzset{external/force remake}

\usepackage{pgfplots}
\pgfplotsset{compat=1.14}

\newcommand{\mean}[1]{\left\langle #1 \right\rangle}
\newcommand{\meann}[1]{\langle #1 \rangle}
\newcommand{\ket}[1]{| #1 \rangle}

\DeclareMathOperator\Tr{Tr}

\newcommand{\Li}{\operatorname{Li}}

\begin{document}
\author{Nahuel Freitas}
\affiliation{Complex Systems and Statistical Mechanics, Department of Physics and Materials Science,
University of Luxembourg, L-1511 Luxembourg, Luxembourg}

\author{Massimiliano Esposito}
\affiliation{Complex Systems and Statistical Mechanics, Department of Physics and Materials Science,
University of Luxembourg, L-1511 Luxembourg, Luxembourg}

\title{Information flows in macroscopic Maxwell's demons}

\date{\today}

\begin{abstract}
A CMOS-based implementation of an autonomous Maxwell's demon was recently proposed (Phys. Rev. Lett. 129, 120602) to demonstrate that a Maxwell demon can still work at macroscopic scales, provided that its power supply is scaled appropriately. Here, we first provide a full analytical characterization of the non-autonomous version of that model. We then study system-demon information flows within generic autonomous bipartite setups displaying a macroscopic limit. By doing so, we can study the thermodynamic efficiency of both the measurement and the feedback process performed by the demon. We find that the information flow is an intensive quantity and that, as a consequence, any Maxwell's demon is bound to stop working above a finite scale if all parameters but the scale are fixed. However, this can be prevented by appropriately scaling the thermodynamic forces. These general results are applied to the autonomous CMOS-based demon.
\end{abstract}

\maketitle

\section{Introduction}

A Maxwell's demon is an agent or mechanism able to make a system behave in a way that seems to contradict the second law of thermodynamics \cite{leff2002}. It achieves that by gaining information about the system via measurements, based on which it performs a feedback on the system. In an ideal situation, the feedback control must not provide any energy to the system \cite{Freitas2021pre}. Maxwell's demon models come in two flavors: non-autonomous ones where the feedback is modeled as a parametric time-dependence on the system's dynamics, and autonomous ones, where the mechanism of the demon coupled to the system is explicitly modeled and the system-demon dynamics is homogeneous in time.       

Many implementations of these ideas have been provided over recent years \cite{Strasberg2013, Serreli2007, Toyabe2010, saha2021, Koski2014Jul, Koski2014Sep,Chida2017, Camati2016, Vidrighin2016, Cottet2017, Masuyama2018, Naghiloo2018, Ribezzi-Crivellari2019, Najera-Santos2020, Amano2022}. An electronic implementation of an autonomous Maxwell's demon fully based on CMOS technology (the same used to build regular computers) was recently proposed in \cite{Freitas2022demon}. Its distinctive feature is that it can be made to work at different physical scales. This scale, $\Omega$, is controlled by the size of the MOS transistors involved in the proposed circuit. It was shown that when all the other parameters are fixed, the rectification effects associated with the action of the demon disappear above a finite scale $\Omega^*$. However, when the powering available to the demon to perform its functions is appropriately scaled with $\Omega$, the demon can in principle continue to work at any scale. In that case, the price to pay is a decreasing thermodynamic efficiency, that was shown to scale as $\eta \propto 1/\Omega^2\log(\Omega)$.

In Section \ref{sec:non-auto} we start by studying a non-autonomous version of the CMOS-demon proposed in \cite{Freitas2022demon}. This enables an exact analytical treatment of the rectification and captures the essential features of the demon mechanism. In Section \ref{sec:auto} we revisit (in more mathematical detail) the autonomous implementation proposed in \cite{Freitas2022demon} and compare the results with those of the non-autonomous version. 
In Section \ref{sec:inf_flow} we turn to the most important part of the paper: the general analysis of autonomous Maxwell demons displaying a macroscopic limit in terms of the information thermodynamics theory developed in \cite{horowitz2014, hartich2014}. This allows us to study the information flows between the demon and the system and to define independent thermodynamic efficiencies for the measurement and feedback processes performed by the demon. We derive general scaling laws for the information flows and the thermodynamic efficiencies. In particular we find that if all extra parameters are fixed, the information flow is scale invariant for any autonomous bipartite system with a macroscopic limit and as a consequence any autonomous Maxwell's demon will stop operating above a finite scale. Scaling power is thus essential to produce macroscopic Maxwell demons.

\section{Non-autonomous CMOS-demon}
\label{sec:non-auto}

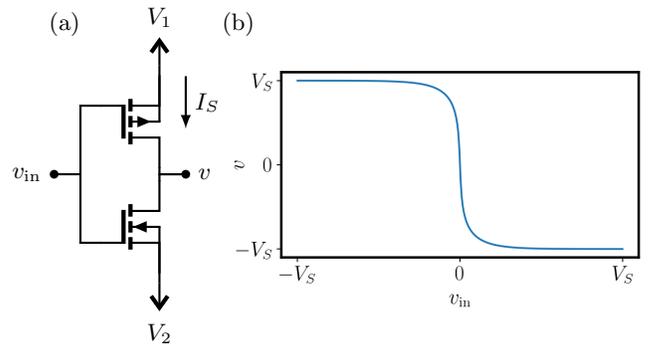
\begin{figure}
\begin{minipage}{.16\textwidth}
\ctikzset{bipoles/length=1.1cm}
\begin{circuitikz}[scale=.7, thick]
\draw (0,1) node[pigfete] (pfet) {};
\draw (0,-1) node[nigfete] (nfet) {};
\draw (pfet.S) -- ++(0,.2) node[vcc]{$V_1$};
\draw (nfet.S) -- ++(0,-.2) node[vee]{$V_2$};
\draw (nfet.G) -| (-1.5,0);
\draw (pfet.G) -| (-1.5,0);
\draw (-1.5,0) -- ++(-.5,0) node[circ, label=left:$v_\text{in}$]{};
\draw (nfet.D) |- (.5,0);
\draw (pfet.D) |- (.5,0) node[circ, label=right:$v$]{};
\draw [-latex] (pfet.S) ++(.5,0) -- ++(0,-1) node[midway,right]{$I_S$};
\draw (-1.8,2.85) node[]{\small (a)};
\draw (1.5,2.85) node[]{\small (b)};
%
\end{circuitikz}
\end{minipage}
\begin{minipage}{.29\textwidth}
\includegraphics[scale=.168]{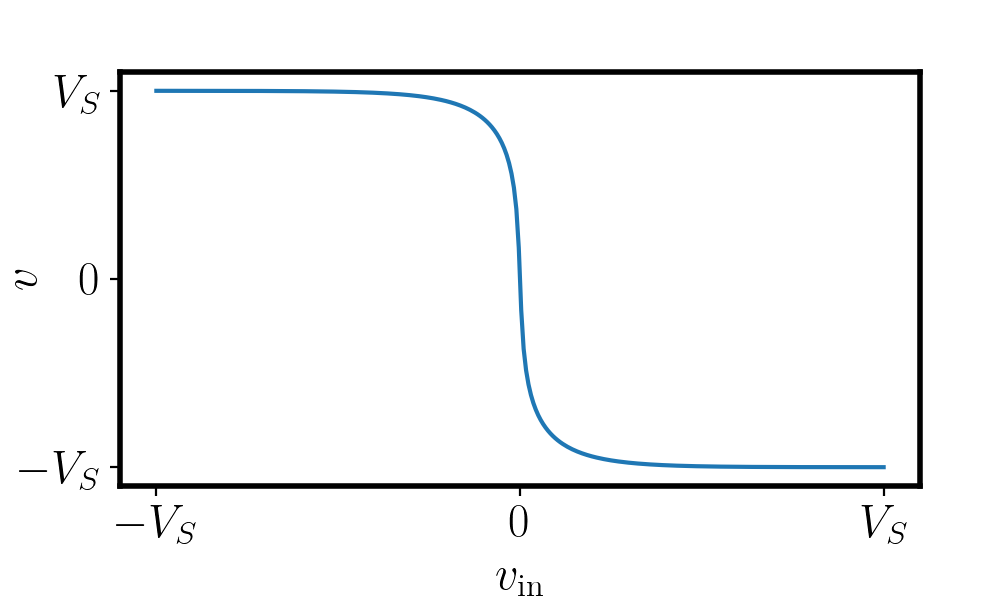}
\end{minipage}
\caption{ (a) Common implementation of a NOT gate with CMOS technology.
(b) Deterministic output voltage as a function of the input (for $V_\text{1}=-V_\text{2} = \Delta V_S/2>0$).}
\label{fig:inverter_diagram}
\end{figure}

\begin{figure}
\centering
\includegraphics[scale=.58]{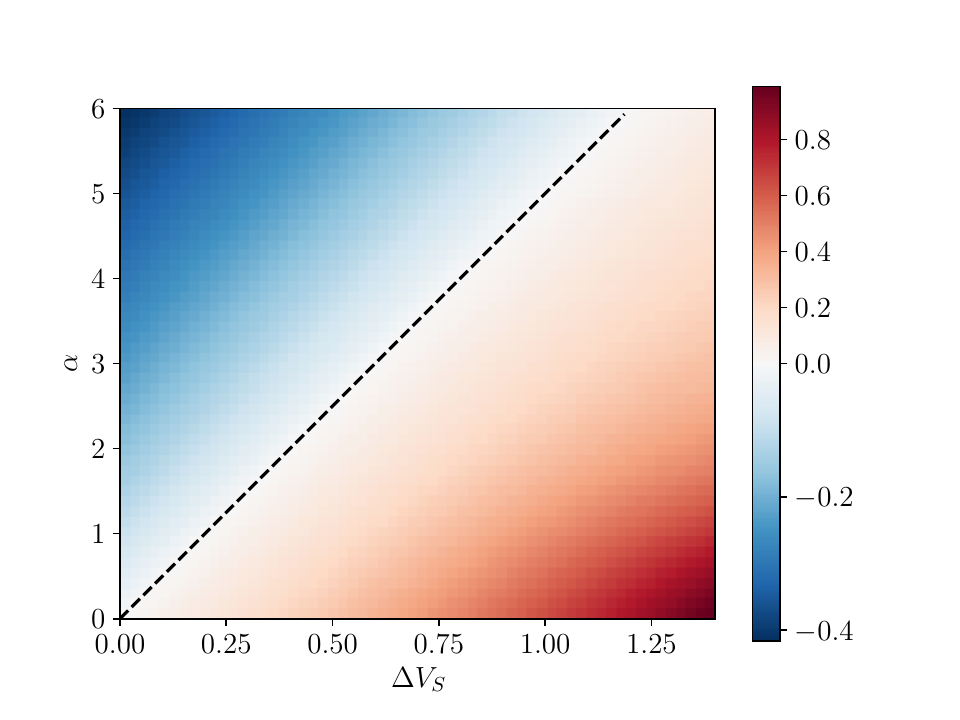}
\caption{Average electric current $\mean{I_S}$ at steady state for the feedback-controlled CMOS inverter as a function of the voltage bias $\Delta V_S$ and the feedback parameter $\alpha$ ($v_e = 0.1$, $n=1$). Negative values (blue) correspond to a current flowing against the voltage bias, or equivalently in the upward direction. The dashed line marks the boundary between the two phases, and is given by Eq. \eqref{eq:current_inv_approx}.}
\label{fig:nonauto_current}
\end{figure}

In this section we show how a simple feedback control protocol in the CMOS inverter can make the electrical current through it flow against the powering bias. As shown in Figure \ref{fig:inverter_diagram}, a CMOS inverter is composed of an 
nMOS transistor and a pMOS transistor. 
We will consider the sub-threshold operation of these devices \cite{wang2006}.
In a three-terminal nMOS transistor, electrical conduction between gate and
source terminals is enhanced when the gate-source voltage is positive, and
reduced when that voltage is negative.  This relation is reversed for pMOS
transistors: conduction is enhanced (reduced) for negative (positive)
gate-source voltages. As a consequence, when a voltage bias is applied by
connecting the drain terminals to voltage sources $V_1=-V_2=\Delta V_S/2>0$, the
input-output transfer function of the inverter has the typical shape shown in
Figure \ref{fig:inverter_diagram}-(b). For positive values of the input, conduction through the nMOS transistor dominates and the output voltage approaches $V_2$. The situation is reversed for negative input voltages, and the output voltage approaches $V_1$.
Let us first consider the situation where no bias is applied: $V_1=V_2=0$. In that case the circuit will attain thermal equilibrium, and the steady state fluctuations
in the output voltage $v$ will be given by the Gibbs distribution
$P_\text{eq}(v) \propto e^{-\beta U(v)}$, where $\beta=1/k_b T$ is the inverse
temperature, and $U(v) = v^2/2C$ is the electrostatic energy for a given value
of voltage ($C$ is the output capacitance of the inverter). 
In the following, all voltages will be expressed in units of the thermal voltage $V_T = k_bT/q_e$, where $q_e$ is the positive electron charge.
Lets now consider the following feedback protocol: at time $t$ the output voltage $v$ is measured and a voltage $v_\text{in} = -\alpha v$ is applied to the input, with $\alpha >0$.
Thus, if a positive fluctuation $v>0$ is observed at time $t$, at subsequent
times conduction through the upper pMOS transistor will be enhanced with
respect to conduction through the nMOS transistor (since the input voltage at
the gates will be negative). Therefore, the excess charge will be most likely
dissipated through the pMOS transistor, generating a net upward current. In a similar way, when a negative fluctuation $v<0$ is observed, it will most probably be compensated by conduction events through the bottom nMOS transistor, also generating a net upward current. Thus, by the repeated application of the feedback protocol, charge can be made to flow in the upward direction, in absence of a voltage bias. A net upward electric current could also be observed even if a small downward bias $V_1 - V_2 = \Delta V_S > 0$ is applied.  
This is indeed confirmed by numerical results shown in Figure \ref{fig:nonauto_current}.
We see that as we increase the biasing voltage $V_S$, a higher amplification factor $\alpha$ is required by the feedback protocol in order to reverse the direction of the current. When that happens, the entropy production rate 
$\dot \Sigma = \Delta V_S \mean{I_S}/T$ is actually negative ($\mean{I_S}$ is the average steady state electric current through the inverter, and therefore $\Delta V_S \mean{I_S}$ is the rate of heat dissipation). According to the second law of thermodynamics, and the modern understanding of Maxwell's demons and information engines, that negative entropy production must be compensated by a positive entropy production in the system implementing the feedback control protocol. This was studied in the autonomous implementation
proposed in \cite{Freitas2022demon}, that is detailed in the next section. Before that, we review the deterministic description of the CMOS inverter in sub-threshold operation, and also its stochastic counterpart leading to the exact results in Figure \ref{fig:nonauto_current}. 

\subsection{Deterministic dynamics}

The deterministic circuit analysis of the inverter in Figure \ref{fig:inverter_diagram}-(a) leads to the following dynamical equation for the output voltage $v(t)$:
\begin{equation}
    C d_t v(t) = I_\text{p}(v(t), v_\text{in};\Delta V_S) - I_\text{n}(v(t), v_\text{in};\Delta V_S) 
    \label{eq:det_dyn}
\end{equation}
where $I_\text{n/p}(v,v_\text{in})$ is the current through the n/p transistor for given values of $v$ and $v_\text{in}$, and $C$ is the output capacitance of the inverter (or the effective capacitance including the one of the next stage to which the inverter is connected to). 
In the sub-threshold regime of operation, the currents are given by:
\begin{equation}
 I_\text{p}(v, v_\text{in};\Delta V_S)=I_0e^{(\Delta V_S/2-v_\text{in}-V_\text{th})/n}\!\left(1\!-\!e^{-(\Delta V_S/2-v)}\right)
 \label{eq:iv_pmos}
\end{equation}
and $I_\text{n}(v, v_\text{in};\Delta V) = I_\text{p}(-v, -v_\text{in};\Delta V)$. In the previous expression, $V_\text{th}$ (the threshold voltage) and $n\geq 1$ (the slope factor), are phenomenological parameters characterizing the transistors.
The deterministic output voltage $v^*$ for a given $v_\text{in}$ are found by solving $d_t v(t) = 0$, 
or equivalently $I_\text{p}(v^*, v_\text{in}) = I_\text{n}(v^*, v_\text{in})$. An explicit expression for $v^*(v_\text{in})$ can be obtained from (for $n=1$):
\begin{equation}
\begin{split}
e^{v^*- v_\text{in}} =
\sqrt{1 \!+\! e^{\Delta V_S} (\cosh(v_\text{in})-1)/2} - e^{\Delta V_S/2} \sinh(v_\text{in})
\end{split}
\end{equation}

In general, we always have $V_1 > v^* > V_2$ and $v^* \to 0$ for $v_\text{in} \to 0$. From the first property it follows that both currents are positive for any value of $v_\text{in}$. Then, no inversion of the current is possible at the deterministic level. The second property justifies to perform a series expansion in both $v$ and $v_\text{in}$ for $v_{in} \ll 1$. Doing that, it is easy to see that the gain for small inputs is:
\begin{equation}
    \alpha_S = \frac{\partial v^*}{\partial v_\text{in}}\bigg\rvert_{v_\text{in}=0} = 
    \left(1-e^{\Delta V_S/2}\right)/n 
    \label{eq:gain_inv}
\end{equation}
Also, we note that $v^*$ is independent of the parameters $I_0$ and $V_\text{th}$, since they only fix the global timescale 
\begin{equation}
\tau_0 = (I_0 e^{-V_\text{th}/n}/q_e)^{-1},
\end{equation}
and do not affect the steady state solution. 

The deterministic output voltage $v^*$ is always unique for a given fixed value of $v_\text{in}$. This is not anymore the case for the feedback protocol mentioned above. We assume that the repeated application of the protocol (that is, the measurement of the output voltage $v$ and the subsequent adjustment of the input voltage as $v_\text{in} = -\alpha v$)
is sufficiently fast compared to the timescale $\tau_0$.
Then the steady state output voltage can be obtained by solving $I_\text{p}(v^*, -\alpha v^*;\Delta V_S) = I_\text{n}(v^*, -\alpha v^*; \Delta V_S)$. 
This equation has $v^*=0$ as the single solution for 
$\alpha \leq \alpha_c \equiv n/\left(e^{\Delta V_S/2}-1\right)$. For $\alpha > \alpha_c$, two additional symmetric solutions $v^*_\pm$ appear and the solution $v^*=0$ becomes unstable. Thus, under the feedback control protocol, the system becomes bistable when the amplification factor is above the critical value $\alpha_c$. Note that the condition $\alpha > \alpha_c$ can be interpreted in the following way: the product of the inverter gain $|\alpha_S|$ in Eq. \eqref{eq:gain_inv} and the amplification factor $\alpha$ of the feedback protocol must be greater than 1.
The electric current is also positive in any of the solutions of the bistable phase, so rectification is still not possible at the deterministic level. 

\subsection{Stochastic dynamics}

Conduction through the MOS transistors is not deterministic. The number of charges transported through the channel of a MOS transistor during a given time interval is a stochastic quantity that, in the sub-threshold regime of operation, displays shot noise \cite{sarpeshkar1993}. This can be described by an effective model that assigns two Poisson processes to every transistor, corresponding to elementary conduction events in the forward and backward directions \cite{freitas2021}. The rates that must be assigned to the two Poisson processes can be computed from the I-V curve characterization of the device and the requirement of thermodynamic consistency, as explained in \cite{freitas2021}. Then, the stochastic dynamics of a given circuit can be mapped to a Markov jump process in the space of possible circuit states, and the probability distribution over those states evolves according to a Markovian master equation. For the CMOS inverter, the master equation reads:
\begin{equation}
d_t P(v, t) = \sum_{\rho}\lambda_\rho(v-\Delta_\rho v_e)P(v-\Delta_\rho v_e,t)-\lambda_\rho(v)P(v,t),
\label{eq:master_eq_inv}
\end{equation}
where $v$ is the output voltage (that can only take discrete values spaced by the elementary voltage $v_e = q_e/C$), and 
the index $\rho$ runs over all the possible jump processes,
with Poisson rates $\lambda_{\pm}^\text{n/p}(v,v_\text{in})$.
For example, the rates $\lambda^\text{p}_{\pm}(v,v_\text{in})$ correspond to conduction events through the pMOS transistor, for which $v \to v \pm v_e$, while the rates $\lambda^\text{n}_{\pm}(v,v_\text{in})$ are associated conduction events through the nMOS transistor,  where $v \to v \mp v_e$. The numbers $\Delta_\rho$ indicate the change in $v/v_e$ for each jump, and therefore we have $\Delta_\pm^\text{p} = \pm 1$ and $\Delta_\pm^\text{n} = \mp 1$. 

In order to ensure thermodynamic consistency, the Poisson rates must satisfy the local detailed balance (LDB) condition, \begin{equation}
    \log \frac{\lambda_\rho(v)}{\lambda_{-\rho}(v+\Delta_\rho v_e)} = \sigma_\rho/k_b,
    \label{eq:ldb}
\end{equation}
according to which the log-ratio of forward and backward rates corresponding to a given 
transition equals the entropy production $\sigma_\rho$ during that transition. For this kind of 
electronic circuits, the entropy production can be computed as $\sigma_\rho = -Q_\rho/T$, where 
$Q_\rho = U(v+\Delta_\rho v_e) - U(v) + W_\rho$ is the heat provided by the thermal environment 
during the transition ($U(v)$ is the electrostatic energy and $W_\rho$ the non-conservative 
work performed by the voltage sources during the transition). The Poisson rates 
$\lambda_\pm^\text{n/p}(v, v_\text{in})$ can be determined from the I-V curve characterization 
of the transistors in Eq. \eqref{eq:iv_pmos} and the LDB condition in Eq. \eqref{eq:ldb}. For 
the pMOS transistor, they read:
\begin{equation}
\begin{split}
\lambda_+^\text{p}(v,v_\text{in}) &= \tau_0^{-1} \: e^{(\Delta V_S/2-v_\text{in})/\text{n}} \qquad\\
\lambda_-^\text{p}(v,v_\text{in}) &= \lambda_+^p(v,v_\text{in}) \: e^{-(\Delta V_S/2-v)} \: e^{-v_e/2},
\end{split}
\label{eq:mos_rates}
\end{equation}
while for the nMOS transistor we have $\lambda_\pm^\text{n}(v,v_\text{in}) = \lambda_\pm^\text{p}(-v,-v_\text{in})$. 

We are interested in obtaining the steady state solution of Eq. \eqref{eq:master_eq_inv}. Since this is a one-dimensional problem, the steady state distribution is fully characterized by the condition that the total rate of transitions $v\to v+v_e$ must balance the total rate of the reverse transitions $v+v_e \to v$. This condition can be expressed as the recurrence relation
\begin{equation}
    P_\text{ss}(v) = \frac{\lambda_+^\text{p}(v-v_e,v_\text{in})+\lambda_-^\text{n}(v-v_e,v_\text{in})}{\lambda_-^\text{p}(v,v_\text{in})+\lambda_+^\text{n}(v,v_\text{in})} P_\text{ss}(v-v_e),
    \label{eq:recursive_ss}
\end{equation}
that can be solved to find $P_\text{ss}(v)$.
So far we have considered the input voltage $v_\text{in}$ to be fixed. The steady state $P^\text{fb}_\text{ss}(v)$ attained under the action of the idealized feedback protocol described above can be found by just replacing $v_\text{in} \to -\alpha v$ in the previous expressions. 

Once the steady state is determined, the average values of the net electric current through both transistors can be computed
as
\begin{equation}
    \mean{I_\text{n/p}} = q_e \sum_{v} P^\text{fb}_\text{ss}(v) \left(\lambda_+^\text{n/p}(v,-\alpha v)-\lambda_-^\text{n/p}(v,-\alpha v)\right).
    \label{eq:mean_currents}
\end{equation}
Since the two transistors are connected in series, at steady state we have $\mean{I_\text{n}} = \mean{I_\text{p}} \equiv \mean{I_S}$. This is how the results in Figure \ref{fig:nonauto_current} were obtained.

\subsection{Macroscopic limit}

We will now study what is the stochastic behaviour of the CMOS inverter at different scales. For this, we will analyze how the parameters entering the problem are modified as the physical dimensions of the transistors are increased. There are two relevant length scales: the width $W$ and the length $L$ of the conduction channel in each transistor \cite{tsividis1987}. For fixed $L$, both the parameter $I_0$ appearing in Eq. \eqref{eq:iv_pmos} (that enter the Poisson rates through the timescale $\tau_0$) and the output capacitance $C$ are proportional to $W$. Thus, the Poisson rates $\lambda_\pm^\text{n/p}(v)$ increase as $W$, while the elementary voltage $v_e = q_e/C$ decreases as $W^{-1}$. 
From these two observations it follows that the probability distribution $P(v,t)$ satisfies a large deviations (LD) principle in the limit $W \to \infty$ \cite{Gopal2022, Freitas2022ESL}. This means that fluctuations away from the deterministic behaviour are exponentially suppressed in $W$. Taking $\Omega \equiv v_e^{-1} \propto W$ as a scale parameter, the LD property is expressed mathematically as the existence of the limit
$f(v,t) = \lim_{v_e \to 0} -v_e\log(P(v,t))$, or equivalently:
\begin{equation}
P(v,t) \underset{v_e \to 0}{\asymp} e^{-(f(v,t) + o(v_e))/v_e },
\label{eq:lda}
\end{equation}
where $f(v,t)$ is the rate function (it gives the rate at which the probability of fluctuation $v$ decreases with $v_e^{-1}$). Indeed, plugging the previous ansatz in the master equation \eqref{eq:master_eq_inv} and keeping only 
the dominant terms in $v_e \to 0$, we find that the rate function must satisfy the evolution equation:
\begin{eqnarray}
 d_t f(v,t) =\sum_\rho \omega_\rho(v)\left[1-e^{\Delta_\rho d_v f(v,t)}\right].
 \label{eq:rf_dyn}
\end{eqnarray}
where $\omega_\rho(v) = \lim_{v_e \to 0} v_e \lambda_\rho(v)$ are the scaled Poisson rates. The minimum $v(t)$ of the rate function $f(\cdot, t)$ at time $t$ is the most probable value, and it can be seen that it evolves according to the closed deterministic dynamics of Eq. \eqref{eq:det_dyn}. In this way we see how the deterministic dynamics emerges from the stochastic one. 

From Eq. \eqref{eq:rf_dyn}, the steady state rate function $f_\text{ss}(v)$ satisfies:
\begin{equation}
    e^{-d_v f_\text{ss}(v)} = \frac{\omega_+^\text{p}(v,v_\text{in})+\omega_-^\text{n}(v,v_\text{in})}{\omega_-^\text{p}(v,v_\text{in})+\omega_+^\text{n}(v,v_\text{in})}.
\end{equation}
Note that this equation can also be obtained as 
the ${v_e \to 0}$ limit of the recursive relation 
in Eq. \eqref{eq:recursive_ss}. Again, the steady state rate function $f^\text{fb}_\text{ss}(v)$ corresponding to the application of the feedback control protocol satisfies a similar equation where $v_\text{in}$ is replaced by $-\alpha v$. Solving for $d_v f^\text{fb}_\text{ss}(v)$ and  integrating in $v$, it is possible to obtain the following explicit expression (up to an arbitrary constant):
\begin{equation}
    f_\text{ss}^\text{fb}(v) = \frac{v^2 + v \Delta V_s}{2} + \frac{n}{n+2\alpha} \left[ L(v, \Delta V_S) - L(v, \Delta V_S) \right]
\end{equation}
where $L(v,\Delta V) = \Li_2\left(-\exp(\Delta V/2 + v(1+2\alpha/n))\right)$, and $\Li_2(\cdot)$ is the polylogarithm function of second order. 

We can use the previous result to compute the average currents according to Eq. \eqref{eq:mean_currents}, which for $v_e \ll 1$ can be approximated as:
\begin{equation}
    \mean{I_S} \simeq \!\int\! dv \: \frac{e^{-v_e^{-1} f^\text{fb}_\text{ss}(v)}}{Z_\text{ss}} \underbrace{q_e \left(\lambda_+^\text{p}(v,-\alpha v)-\lambda_-^\text{p}(v,-\alpha v)\right)}_{I(v)},
    \label{eq:macro_average_current}
\end{equation}
where $Z_\text{ss} = \int dv \exp(-v_e^{-1} f^\text{fb}_\text{ss}(v))$. For $v_e \ll 1$ only small fluctuations around the minimum of the rate function will contribute to the integral, and for $\alpha < \alpha_c$ that minimum is $v=0$. Expanding the function $I(v)$ defined above around $v=0$, we obtain (using the fact that the distribution of $v$ is symmetric around $v=0$):
\begin{equation}
    \mean{I_S} \simeq I(0) + I^{\prime\prime}(0) \mean{v^2}/2.
    \label{eq:current_exp}
\end{equation}
Also, a Gaussian approximation to the steady state for $\alpha<\alpha_c$ can be obtained by expanding the rate function to second order in $v$. That expansion reads (again up to a constant):
\begin{equation}
    f_\text{ss}^\text{fb}(v) =  \frac{1-\alpha/\alpha_c}{1+e^{V_S}} v^2 + \mathcal{O}(v^3).
\end{equation}
Therefore, the steady state voltage variance is 
\begin{equation}
\mean{v^2} \simeq \frac{v_e \left(e^{\Delta V_S/2}+1 \right)}{2(1-\alpha/\alpha_c)}.
\label{eq:var_inv_fb}
\end{equation}
From Eqs. \eqref{eq:var_inv_fb}, \eqref{eq:current_exp} and the definition of $I(v)$ in 
Eq. \eqref{eq:macro_average_current} one can obtain an exact expression for the steady state current, valid to first order in $v_e$, which is however not very simple. To lower order in $\Delta V_S$ it reduces to (neglecting terms proportional to $\Delta V_S v_e$)
\begin{equation}
    \mean{I_S} \simeq \frac{q_e}{\tau_0}\left(
    \Delta V_S/2 - \alpha v_e/n
    \right)
    .
    \label{eq:current_inv_approx}
\end{equation}
Thus, the minimum value of the amplification factor 
needed for the average current to flow against the 
applied bias is $\alpha_m = n\Delta V_S/2v_e$. We note 
that $\alpha_m \to \infty$ in 
the macroscopic limit $v_e \to 0$. Thus, to reverse the 
current is increasingly hard as one goes deeper in the 
macro limit. This is natural, since the rectification 
strategy exploits the fluctuations in the output voltage, and these fluctuations are negligible in the macro limit (their variance scales as $v_e$, see Eq. \eqref{eq:var_inv_fb}). As we will see in the next section, the demon or agent implementing the feedback protocol must invest an increasing amount of energy in order to achieve higher amplification factors. 

\section{Autonomous CMOS-demon}
\label{sec:auto}

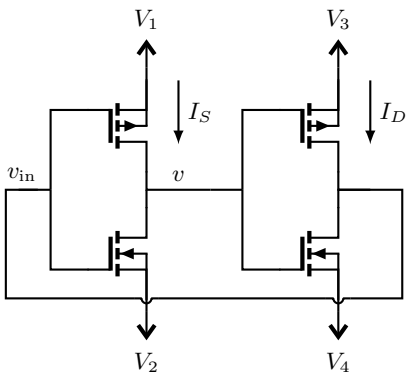
\begin{figure}
\centering
\ctikzset{bipoles/length=1.1cm}
\ctikzset{bipoles/crossing/size=0.3}
\begin{circuitikz}[scale=.85, thick]
\draw (0,1) node[pigfete] (pfet1) {};
\draw (0,-1) node[nigfete] (nfet1) {};
\draw (pfet1.S) -- ++(0,.2) node[vcc]{$V_1$};
\draw (nfet1.S) -- ++(0,-.2) node[vee]{$V_2$};
\draw (nfet1.G) -| (-1.5,0);
\draw (pfet1.G) -| (-1.5,0);
\draw (-1.5,0) -- ++(-.5,0);
\draw (nfet1.D) -- (0,0);
\draw (pfet1.D) |- (.5,0) node[above]{$v$} -- ++(.5,0);
\draw [-latex] (pfet1.S) ++(.5,0) -- ++(0,-1) node[midway,right]{$I_S$};
\begin{scope}[xshift=3cm, xscale=1]
\draw (0,1) node[pigfete, xscale=1] (pfet2) {};
\draw (0,-1) node[nigfete, xscale=1] (nfet2) {};
\draw (pfet2.S) -- ++(0,.2) node[vcc]{$V_3$};
\draw (nfet2.S) -- ++(0,-.2) node[vee]{$V_4$};
\draw (nfet2.G) -| (-1.5,0);
\draw (pfet2.G) -| (-1.5,0);
\draw [-latex] (pfet2.S) ++(.5,0) -- ++(0,-1) node[midway,right]{$I_D$};
\draw (-1.5,0) -- ++(-.5,0);
\draw (nfet2.D) |- (.5,0) -- ++(.5,0) -- ++(0,-1.7)  to[xing] ++(-2,0) -- ++(-1,0) to[xing] ++(-2,0) -- ++(-1.2,0) -- ++(0,1.7) -- ++(.5,0) node[midway, above]{$v_\text{in}$};
\draw (pfet2.D) |- (.5,0);
\end{scope}
\end{circuitikz}
\caption{CMOS implementation of an autonomous Maxwell's demon. If we break the left/right symmetry of the circuit by choosing powering biases $V_3 - V_4 > V_1 -V_2$, we can consider that the right inverter monitors the thermal fluctuations in the output of the left inverter, and by appropriately acting back on its input, can make the electric current $I_S$ to flow against the bias $V_1-V_2$. For simplicity, all transistors are assumed to have identical parameters.}
\label{fig:auto_circuit}
\end{figure}

\begin{figure}
\centering
\includegraphics[scale=.58]{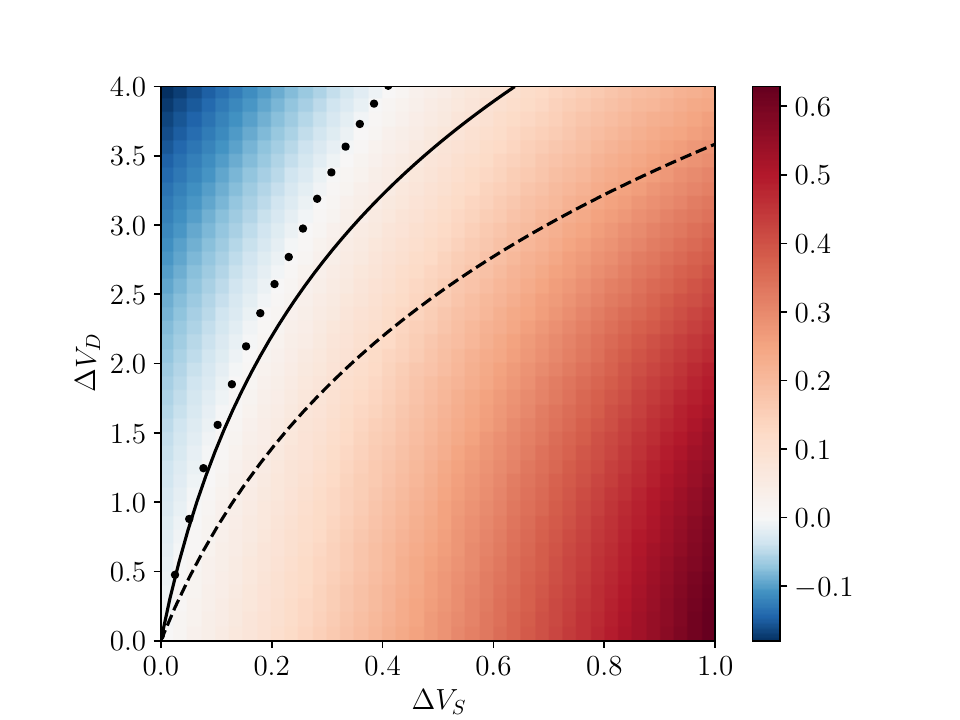}
\caption{Average electric current $\mean{I_S}$ at steady state for the autonomous circuit of Figure \ref{fig:auto_circuit} as a function of the voltage biases $\Delta V_S$ and $\Delta V_D$ ($v_e=0.1$, $n=1$). Negative values (blue) correspond to a current flowing against the voltage bias of the first inverter. The dotted line indicates the actual boundary between positive and negative currents. The solid black line shows the approximation given by Eq. \eqref{eq:vdmin}, while the dashed black line shows the estimation of Eq. \eqref{eq:vdmin_na} based on the non-autonomous model.} 
\label{fig:auto_current}
\end{figure}

A feedback protocol similar to the non-autonomous one can be implemented using an additional inverter (see Figure \ref{fig:auto_circuit}), which plays the role of a Maxwell's demon monitoring the state of a system
(the output of the first inverter) and performing some feedback on it (by
changing the input voltage). The second inverter is powered by voltages
$V_3=-V_4 = \Delta V_D/2$. Thus, the bias $\Delta V_D$ controls the power available to the
demon. We consider for simplicity that the two transistors of the first inverter are equivalent to the two transistors of the second inverter. Thus, the only asymmetry between system and demon is introduced by the different biasing voltages $\Delta V_S$ and $\Delta V_D$. As we explain below, it is possible to compute the steady state distribution of the two degrees of freedom of the circuit, $v$ and $v_\text{in}$, 
from which average currents can be obtained. In that way we obtain the results of Figure \ref{fig:auto_circuit}. We see that if the bias $\Delta V_D$ applied to the demon is sufficiently high, then its action is able to reverse the direction of the electric current through the system. Before further analyzing the performance of this autonomous implementation of an electronic Maxwell's demon, we explain how to model the dynamics of the circuit in Figure \ref{fig:auto_circuit}, again both at the deterministic and stochastic levels. It is important to note that this autonomous model automatically takes into account some realistic effects that were not considered in the previous non-autonomous implementation, namely the intrinsic noise affecting the measurement and feedback processes, and the delay between them.

\subsection{Deterministic dynamics}
The deterministic equations
of motion for the voltages $v$ and $v_\text{in}$ are
\begin{equation}
\begin{split}
    C d_t v &= I_\text{p}(v, v_\text{in} ; \Delta V_S) - I_\text{n}(v, v_\text{in} ; \Delta V_S) \\
    C d_t v_\text{in} &= I_\text{p}(v_\text{in}, v ; \Delta V_D) - I_\text{n}(v_\text{in}, v ; \Delta V_D)
\end{split}
\end{equation}
Close to thermal equilibrium (that is, for low biases $\Delta V_S$ and $\Delta V_D$), the previous deterministic dynamics has the unique fixed point $v = v_\text{in} = 0$. However the system becomes bistable when $\alpha_S \alpha_D > 1$, where $\alpha_{S/D} \equiv (1-e^{\Delta V_{S/D}/2})/n$ is the gain of each inverter, given by Eq. \eqref{eq:gain_inv}. The bistability can be exploited to store the value of a bit, and in fact for symmetric powering $\Delta V_S = \Delta V_D = \Delta V$ the circuit in Figure \ref{fig:auto_circuit} is the typical CMOS implementation of SRAM memory cells \cite{Freitas2021pre}. In that case, also assuming $n=1$, the bistability is achieved for $\Delta V> 2\log(2)$.

\subsection{Stochastic dynamics}
\label{sec:auto_stoch_dyn}

%
\begin{figure}[ht]
\centering
\includegraphics[scale=.55]{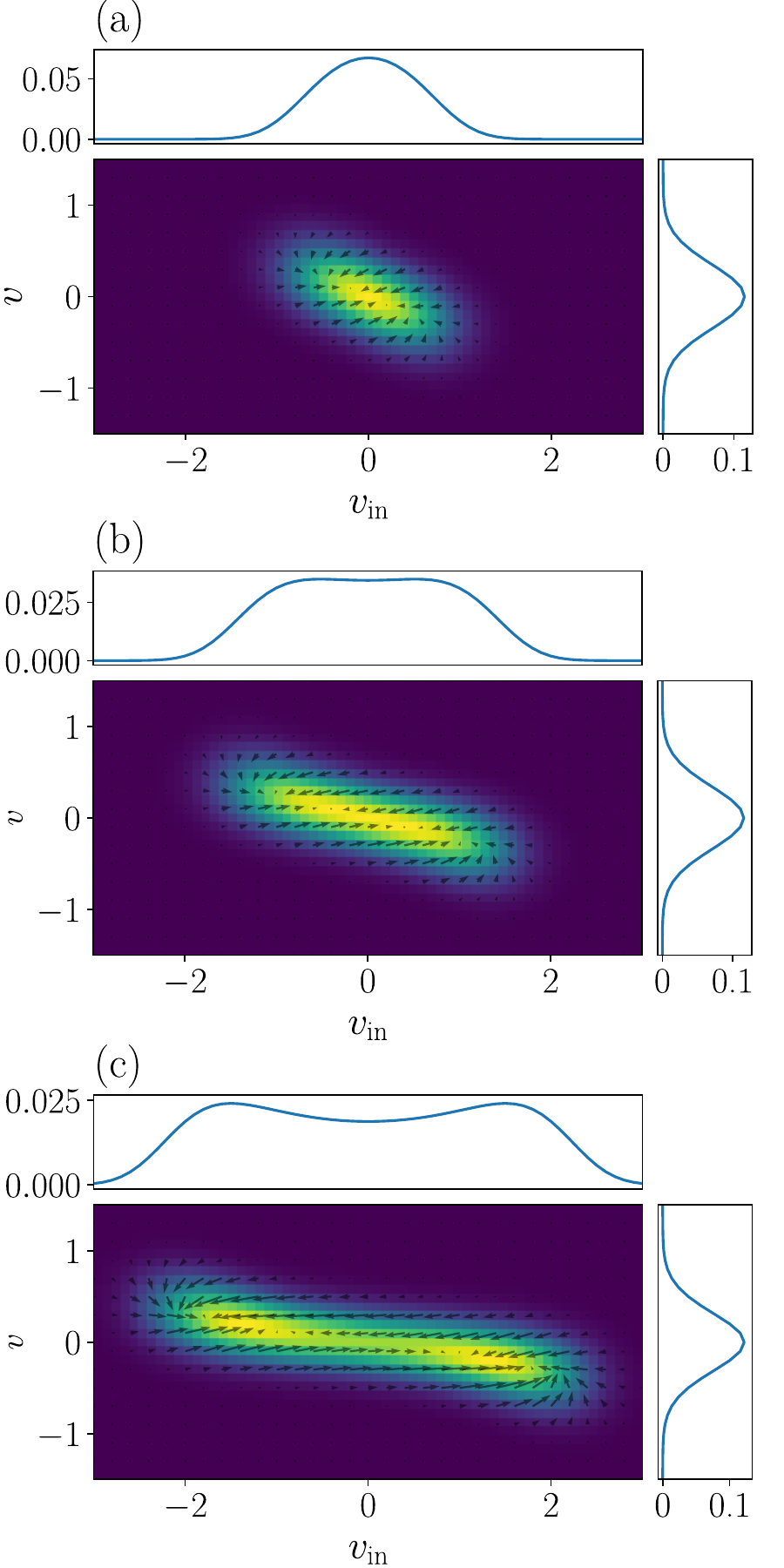}
\caption{Histograms of the steady state distribution $P_\text{ss}(\bm{v})$ for $v_e = 0.1$, $\Delta V_S = 0.4$ and: (a) $\Delta V_D=2$ ($\alpha^2 < 1$, monostable phase), (b) $\Delta V_D\simeq 3.415$ ($\alpha^2 =1$, critical point), 
and (c) $\Delta V_D = 5$ ($\alpha^2 >1$, bistable phase). Lateral and upper plots show the partial distributions for $v$ and $v_\text{in}$, respectively. The arrows indicate the direction and magnitude of the flow $\bm{u}(\bm{v}) = \sum_\rho \bm{\Delta}_\rho \: \omega_\rho(\bm{v}) P_\text{ss}(\bm{v})$ at each point in the state space.}
\label{fig:ss_auto}
\end{figure}

The master equation describing the stochastic evolution of 
the circuit in Figure \ref{fig:auto_circuit} is constructed in the same way as for the simpler implementation of the previous section. It reads
\begin{equation}
d_t P(\bm{v}, t) = \sum_{\rho}\lambda_\rho(\bm{v}-\bm{\Delta}_\rho v_e)P(\bm{v}-\bm{\Delta}_\rho v_e,t)-\lambda_\rho(\bm{v})P(\bm{v},t),
\label{eq:master_eq_auto}
\end{equation}
where now $\bm{v} = (v, v_\text{in})$ is a vector with the two degrees of freedom of the circuit, and the vectors $\bm{\Delta}_\rho$ encode the change in $\bm{v}$ for each jump. For example, for forward transitions through the pMOS and nMOS transistors of the first inverter, we have $\bm{\Delta}_\rho = (1,0)$ and $\bm{\Delta}_\rho = (-1,0)$, respectively. The transition rates $\lambda(\bm{v})$ are obtained in the same way as before (see Eq. \eqref{eq:mos_rates}), and satisfy the LDB conditions ensuring thermodynamic consistency.

In this case, the determination of the steady state $P_\text{ss}(\bm{v})$ is not as straightforward as before. 
Since the voltages $v$ and $v_\text{in}$ take only discrete values spaced by the elementary voltage $v_e$, the steady state can be found numerically by truncating the infinite state space to a finite space with $\max(|v/v_e|,|v_\text{in}/v_e|) \leq  N$, for sufficiently large $N$. Then, the master equation can be written as $d_t \ket{P(t)} = W \ket{P(t)}$, where the probability 
vector $\ket{P(t)}$ has $(2N+1)^2$ components and the generator $W$ is a matrix of dimensions $(2N+1)^2 \times (2N+1)^2$. The steady state vector $\ket{P_\text{ss}}$ is given by the right eigenvector of $W$ with zero eigenvalue. Since the matrix $W$ is sparse, that computation can be done efficiently. Typical steady state distributions for different values of $\alpha^2$ are shown in Figure \ref{fig:ss_auto}. 

The results of the above procedure are shown in Figure \ref{fig:auto_current}. We see that if $V_D/V_S$ is sufficiently high then the current through the first inverter is effectively reversed. The boundary between positive and negative values of that current can be estimated in the following way. In the non-autonomous implementation, a minimum amplification factor $\alpha_m = nV_s/q_e$ was required for the inversion of the current. From Eq. \eqref{eq:gain_inv}, we know that the second inverter can provide that minimum amplification factor if the biasing voltage $V_D$ is above 
\begin{equation}
 \Delta V_D^{m} = 2\log(1+n^2\Delta V_S/2v_e).  
\label{eq:vdmin_na}
\end{equation}
The previous result is plotted with a dashed line in Figure \ref{fig:auto_current}, where we see that it underestimates the minimum value of $V_D$ required for inversion. The reason is that the previous estimation does not consider the intrinsic noise and delay in the demon side. An improved estimation is obtained below.

\subsection{Macroscopic limit}
\label{sec:macro_lim}

The solution to Eq. \eqref{eq:master_eq_auto} also satisfies a LD principle
\begin{equation}
P(\bm{v},t) \underset{v_e\to 0}{\asymp}  e^{-(f(\bm{v},t) + o(v_e))/v_e},
\label{eq:lda_auto}
\end{equation}
where the rate function $f(\bm{v}, t)$ evolves according to 
\begin{eqnarray}
 d_t f(\bm{v},t) =\sum_\rho \omega_\rho(\bm{v})\left[1-e^{\bm{\Delta}_\rho \cdot \nabla f(\bm{v},t)}\right],
 \label{eq:rf_dyn_auto}
\end{eqnarray}
but in this case the steady state rate function $f_\text{ss}(\bm{v})$ cannot be computed exactly. One can still perform a Gaussian approximation. Thus, in the monostable phase where the minimum of the rate function is $\bm{v}=0$, we have:
\begin{equation}
    f_\text{ss}(\bm{v}) = \bm{v}^T \mathcal{C}^{-1} \bm{v}/2 + \mathcal{O}(|\bm{v}|^3),
\end{equation}
where the matrix $\mathcal{C}$ is the solution of 
\begin{equation}
    0 = \mathcal{CA + A^TC + B},
    \label{eq:cov_lyapunov}
\end{equation}
and the matrices $\mathcal{A}$ and $\mathcal{B}$ are given by
\begin{equation}
\{\mathcal{A}\}_{ij} = \sum_\rho \partial_{v_i} \omega_\rho(0) (\bm{\Delta}_\rho)_j
\end{equation}
and
\begin{equation}
\{\mathcal{B}\}_{ij} = \sum_\rho \omega_\rho(0) (\bm{\Delta}_\rho)_i (\bm{\Delta}_\rho)_j.
\end{equation}
Solving Eq. \eqref{eq:cov_lyapunov} one finds the following correlators:
\begin{equation}
\begin{split}
    \mathcal{C}_{11} &= 
    \frac{1+ e^{\Delta V_D/2} + e^{\Delta V_S} - e^{(\Delta V_D+\Delta V_S)/2}}{2(e^{\Delta V_D/2}+e^{\Delta V_S/2}-e^{(\Delta V_D+ \Delta V_S)/2})}\\
    \mathcal{C}_{22} &= 
    \frac{1+ e^{\Delta V_D} + e^{\Delta V_S/2} - e^{(\Delta V_D+\Delta V_S)/2}}{2(e^{\Delta V_D/2}+e^{\Delta V_S/2}-e^{(\Delta V_D+ \Delta V_S)/2})}\\
    \mathcal{C}_{12} = \mathcal{C}_{21} &= 
    \frac{e^{(\Delta V_D+ \Delta V_S)/2} -1}{2(e^{(\Delta V_D+\Delta V_S)/2}-e^{\Delta V_S/2}-e^{\Delta V_D/2})}.\\
\end{split}
\label{eq:correlators}
\end{equation}

As in the previous section, the average steady state current through the first inverter can be computed as 
\begin{equation}
\begin{split}
    \mean{I_S} &\simeq \int\!\!\int d^2\bm{v} \: \frac{e^{-v_e^{-1} f^\text{fb}_\text{ss}(\bm{v})}}{Z_\text{ss}} \: \underbrace{q_e \left(\lambda_+^\text{p}(\bm{v})-\lambda_-^\text{p}(\bm{v})\right)}_{I(\bm{v})}\\
    & \simeq I(0) + \Tr[\bm{H}_I(0) \meann{\bm{v}\bm{v}^T}]/2,
    \label{eq:macro_average_current_auto}
\end{split}
\end{equation}
where $\bm{H}_I(\bm{v})$ is the Hessian of $I(\bm{v})$, and $\meann{\bm{v}\bm{v}^T} = v_e \mathcal{C}$ is the covariance matrix of the Gaussian fluctuations around the fixed point (in the second line we have assumed that the distribution of $\bm{v}$ is symmetric upon the point reflection $\bm{v} \to -\bm{v}$, and therefore all odd moments vanish). We can then obtain an analytical expression for the average current $\mean{I_S}$, valid to lower non-trivial order in $v_e \to 0$, which is however too complicated to show here (see the Supplementary Material in \cite{Freitas2022demon}). To lower order in $V_S$ it reduces to (neglecting terms proportional to $\Delta V_S v_e$)
\begin{equation}
    \mean{I_S} = \frac{q_e}{2\tau_0}\left[\Delta V_S -v_e(e^{\Delta V_D/2}-1)\right].
    \label{eq:current_auto_linear}
\end{equation}
From the previous result it follows that the minimum value of $\Delta V_D$ required for inversion is:
\begin{eqnarray}
 \Delta V_D^{m} = 2\log(1+ \Delta V_S/v_e),
 \label{eq:vdmin}
\end{eqnarray}
that differs from Eq. \eqref{eq:vdmin_na} by a factor $2$ accompanying $\Delta V_S$. This is shown with a solid line in Figure \ref{fig:auto_current}. Alternatively, from Eq. \eqref{eq:current_auto_linear} we can say that inversion is impossible above a maximum scale
\begin{eqnarray}
 {v_e^*}^{-1} = |\alpha_D|/\Delta V_S.
 \label{eq:max_scale}
\end{eqnarray}

Whenever the action of the demon manages to make the current $I_S$ to flow against the voltage bias $\Delta V_S $, heat is extracted from the environment of the first inverter and work is realized on the sources fixing the voltages $V_1$ and $V_2$. In that case the entropy production rate at the system side is $\dot \Sigma_S = \mean{I_S} \Delta V_S < 0$. As we have seen, for that to happen energy needs to be consumed by the second inverter, and therefore the entropy production rate at the demon side is $\dot \Sigma_D = \mean{I_D} \Delta V_D > 0$. We can then define the efficiency of the demon as the ratio:
\begin{equation}
    \eta = - \frac{\dot \Sigma_S}{\dot \Sigma_D}\\
         = - \frac{\Delta V_S}{\Delta V_D} \frac{\mean{I_S}}{\mean{I_D}} \leq 1.
    \label{eq:global_eff}
\end{equation}
The last inequality follows from the fact that the total 
entropy production rate $\dot \Sigma = \dot \Sigma_S + \dot \Sigma_D$ is always positive. 
The thermodynamic efficiency defined in Eq. 
\eqref{eq:global_eff} allows to characterize the 
performance of the demon at a global level, and can be 
evaluated for different parameters using the above results.
However, it is also possible to introduce detailed
thermodynamic efficiencies characterizing the measurement 
and feedback processes independently, providing 
additional insight. Then, the net efficiency in Eq. \eqref{eq:global_eff} is recovered as the product of the detailed efficiencies. To see this, we review in the following the basic theory of information flows as developed in \cite{horowitz2014, hartich2014}, and adapt it to Gaussian states.

\section{Information thermodynamics}
\label{sec:inf_flow}

The thermodynamics of measurement and feedback protocols in bipartite systems can be understood in terms of information flows which, as explained in \cite{horowitz2014, hartich2014}, enter as additional terms in the entropy balance of each subsystem. In this section we compute those information flows for general bipartite systems displaying a macroscopic limit, and then apply it to our autonomous CMOS-demon. For this, in the sake of clarity, we will adopt a slightly different notation. We assume that we are dealing with a system with two degrees of freedom $x$
and $y$ ($v$ and $v_\text{in}$ in our example). The Poisson rates
$\lambda_\rho(x|y)$ associated to jumps $x \to x + \delta x_\rho$ depend on the variable $y$, and in the same way the Poisson rates
$\lambda_\rho(y|x)$ associated to jumps $y \to y + \delta y_\rho$ depend on the variable $x$. In our circuit, this corresponds to the fact that the Poisson rates associated to the transistors in the first inverter (which modify the voltage $v$) depend on the voltage $v_\text{in}$ (this is the feedback process in which the demon controls the system), while the rates of the transistors in the second inverter (which modify $v_\text{in}$) depend on the voltage $v$ (this is the measurement process in which the demon acquires information about the state of the system). The jump sizes $\delta x_\rho$ and $\delta y_\rho$ are $\pm v_e$ in our case, with the sign depending on the transition. 

The mutual information between $x$ and $y$ is defined as 
\begin{equation}
\mathcal{I} = \sum_{x,y} P_t(x,y) \log \frac{P_t(x,y)}{P_t(x)P_t(y)},
\end{equation}
where $P_t(x,y)$ is the 
time-dependent global probability distribution, while 
$P_t(x)$ and $P_t(y)$ are the corresponding reduced 
distributions for $x$ and $y$. The time derivative of the 
mutual information accepts the following decomposition 
\cite{horowitz2014}:
\begin{equation}
d_t \mathcal{I} =  \mathcal{\dot I}^x + \mathcal{\dot I}^y  
\end{equation}
where 
\begin{equation}
\begin{split}
    \mathcal{\dot I}^x &= \sum_{x,y} \sum_{\rho>0} j_\rho(x|y) \log \left( \frac{P_t(y|x+\delta x_\rho)}{P_t(y|x)}\right)\\
    \mathcal{\dot I}^y &= \sum_{x,y} \sum_{\rho>0} j_\rho(y|x) \log \left( \frac{P_t(x|y+\delta y_\rho)}{P_t(x|y)}\right)
\end{split}
\label{eq:inf_flow_exact}
\end{equation}
are the contributions due to jumps $x\to x+\delta x_\rho$
and $y \to y + \delta y_\rho$, respectively. In the previous expressions, $j_\rho(x|y) \equiv \lambda_\rho(x|y) P_t(x,y) - \lambda_{-\rho}(x+\delta x_\rho|y) P_t(x+\delta x_\rho,y)$ is the net current along transition $x,y \to x +\delta x_\rho,y$, and $j_\rho(y|x)$ is defined similarly. Note that at steady state we have $d_t \mathcal{I} = 0$ and therefore $\mathcal{\dot I}^x = - \mathcal{\dot I}^y$. 

We will now evaluate the steady state information flows 
$\mathcal{\dot I}^{x/y}$ in the macroscopic limit. Under a 
Gaussian approximation like the one used previously, the log-ratio of conditional probabilities appearing in the equations above can be written as (for simplicity we assume $\mean{x}=\mean{y}=0$):
\begin{equation}
\begin{split}
    \log\left(\frac{P_t(y|x')}{P_t(y|x)}\right) =& \frac{\mean{xy}}{\mean{xx}\mean{yy}\!-\!\mean{xy}^2}\times \\
    &\left[y\:(x'\!-\!x) \!-\! \frac{\mean{xy}}{\mean{xx}} (x'^2\!-\!x^2)/2\right],
\end{split}
\end{equation}
where the averages are computed for $P_t(x,y)$. 
It follows that the information flow $\mathcal{\dot I}^x$ can be 
rewritten as:
\begin{equation}
    \mathcal{\dot I}^x\:  = \frac{\mean{xy}}{\mean{xx}\mean{yy} - \mean{xy}^2} \left[ \mean{y \: \bar d_t x}  - \frac{\mean{xy}}{\mean{xx}} \: d_t\meann{x^2} /2\right],
    \label{eq:inf_flow_gauss}
\end{equation}
where
\begin{equation}
\begin{split}
    \mean{y \: \bar d_t x} &= \sum_{x,y} \sum_{\rho>0} j_\rho(x|y)
    \: y \: \delta x_\rho\\
    d_t \meann{x^2} &= \sum_{x,y} \sum_{\rho>0} j_\rho(x|y)
    \: (2  x \: \delta x_\rho + \delta x_\rho^2).
\end{split}
\end{equation}
The bar in $\bar d_t$ denotes that it is not a total derivative.
The last term in Eq. \eqref{eq:inf_flow_gauss} vanishes at steady state. Then, at the Gaussian level the steady state information flows read:
\begin{equation}
    \mathcal{\dot I}_\text{ss}^x\:  = \frac{\mean{xy}  \mean{y \: \bar d_t x}}{\mean{xx}\mean{yy} - \mean{xy}^2} 
    \:\:\:
    \text{and}
    \:\:\:
    \mathcal{\dot I}_\text{ss}^y\:  = \frac{\mean{xy}  \mean{x \: \bar d_t y}}{\mean{xx}\mean{yy} - \mean{xy}^2}. 
    \label{eq:inf_flow_gauss_ss}
\end{equation}
Note that $\mathcal{\dot I}_\text{ss}^x + \mathcal{\dot I}_\text{ss}^y \propto \meann{x \:\bar d_t y} + \mean{y \:\bar d_t x} = d_t \meann{xy} = 0$ in steady state conditions, as discussed above.
A calculation similar to the one in the Section \ref{sec:macro_lim} shows that to first non-trivial order in the macroscopic limit:
\begin{equation}
\begin{split}
    \meann{y \:d_t x} &= \mean{xy} \sum_{\rho>0}\left[\partial_x \lambda_\rho(x|y)|_{0,0} - \partial_x \lambda_{-\rho}(x|y)|_{0,0} \right]\delta x_\rho \\
    &+\mean{yy}  \sum_{\rho>0} [\partial_y \lambda_\rho(x|y)|_{0,0} - \partial_y \lambda_{-\rho}(x|y)|_{0,0}]\delta x_\rho 
    \label{eq:ydtx}.
\end{split}
\end{equation}
Recalling that in the macroscopic limit we have the scalings 
$\lambda_\rho \propto \Omega$ and $\delta x_\rho, \mean{xx},\mean{yy},\mean{xy} \propto 1/\Omega$ with respect to a scale parameter $\Omega$, we see that $\mean{y \: d_tx} \propto 1/\Omega$. Therefore, the information flows $\mathcal{\dot I}_\text{ss}^{x/y}$ are scale independent.

For the kind of bipartite systems we are considering, it was shown in \cite{horowitz2014} that when the information flows are taken into account in the entropy balance of each subsystem, then the following extended local second laws hold (we assume isothermal settings at temperature $T$):
\begin{equation}
\begin{split}
    \dot \Sigma_x^i &= d_t S_x + \dot \Sigma_x - k_b \mathcal{\dot I}^x\geq 0\\
    \dot \Sigma_y^i &= d_t S_y + \dot \Sigma_y - k_b \mathcal{\dot I}^y \geq 0,
    \label{eq:local_second_laws}
\end{split}
\end{equation}
where $\dot \Sigma_{x/y}^i$, $S_{x/y}$, $\dot \Sigma_{x/y}$ are, respectively, the total irreversible entropy production, the internal entropy, and the entropy flow into the environment of each subsystem. Since $S_{x/y}$ are the entropies associated to the reduced distributions $P_t(x)$ and $P_t(y)$, we see that by adding the two previous equations we recover the usual global second law $\dot \Sigma^i = \dot \Sigma^i_x + \dot \Sigma^i_y = d_t S + \dot \Sigma$, where $\dot \Sigma = \dot \Sigma_x + \dot \Sigma_y$ and $S = S_x + S_y - \mathcal{I}$ is the entropy of the full distribution $P_t(x,y)$.  

\vspace{.2cm}

\subsection{Efficiency of information creation/consumption}

We will now compute the information flows for the electronic Maxwell demon based on the previous result, taking $x=v$ (the output voltage of the first inverter) and $y=v_\text{in}$ (the output voltage of the second inverter). Thus, $x$ represents the system S and $y$ the demon D. Then, when evaluated at steady state, the inequalities in Eq. \eqref{eq:local_second_laws} reduce to 
\begin{equation}
\begin{split}
    \dot \Sigma_S^i &= \dot \Sigma_S + k_b \mathcal{\dot I}\geq 0\\
    \dot \Sigma_D^i &= \dot \Sigma_D - k_b \mathcal{\dot I} \geq 0,
\end{split}
\label{eq:local_second_laws_ss}
\end{equation}
where $\mathcal{\dot I} = -\mathcal{\dot I}^S_\text{ss} = \mathcal{\dot I}^D_\text{ss}$ is the steady state information flow.
As we will see, $\mathcal{\dot I}$ is positive, which means that 
the dynamics of the demon produces correlations, that are then consumed at the system side. 
This allows the entropy flow $\dot \Sigma_S$ to be negative while still respecting the first inequality in Eq. \eqref{eq:local_second_laws_ss}. 

Evaluating Eq. \eqref{eq:ydtx} for our circuit we obtain:
\begin{equation}
\begin{split}
    \meann{yd_t x} &= \mean{xy} \left[ \partial_x I_S^\text{p}|_{0,0} \: v_e + \partial_x I_S^\text{n}|_{0,0} (-v_e)\right]/q_e \\
    &\; \; \; + 
    \mean{yy} \left[\partial_y I_S^\text{p}|_{0,0} \: v_e + \partial_y I_S^\text{n}|_{0,0} (-v_e)\right]/q_e\\
    & = (2v_e/q_e) \left[\mean{xy} \partial_x I_S^\text{p}|_{0,0} + 
    2 \mean{yy} \partial_y I_S^\text{p}|_{0,0}\right]\\
    & = -\frac{2v_e}{\tau_0} \left[\mean{xy} 
    + \mean{yy} \left(e^{\Delta V_S/2} -1\right)\right]
\end{split}
\label{eq:ydtx_eval}
\end{equation}
where $I_S^\text{n/p}(x,y) =  q_e(\lambda_+^\text{n/p}(x,y)-\lambda_-^\text{n/p}(x,y))$
is the electric current through the (p/n)MOS transistor in the first inverter, and in the second line we have used the fact that $\partial_{x/y}I_S^\text{p}|_{0,0} = -\partial_{x/y}I_S^\text{n}|_{0,0}$. Combining Eqs. \eqref{eq:inf_flow_gauss_ss}, \eqref{eq:ydtx_eval}, and the expression for the correlators in Eq \eqref{eq:correlators}, 
we find that the information flow is given by:
\begin{widetext}
\begin{equation}
    \mathcal{\dot I} = \frac{2v_e}{\tau_0} \frac{\left(1-e^{(\Delta V_S + \Delta V_D)/2}\right)\left(e^{\Delta V_S/2}-e^{\Delta V_D/2}\right)}
    {1+e^{\Delta V_S/2}+e^{\Delta V_D/2}+e^{\Delta V_S}+e^{\Delta V_D}-e^{(\Delta V_S + \Delta V_D)/2}} + \mathcal{O}(v_e^{2})
    \label{eq:inf_flow}
\end{equation}
\end{widetext}
which is indeed scale invariant (recall that $\tau_0 \propto v_e \propto 1/\Omega$), as discussed above. 

As already mentioned, the information flow $\mathcal{\dot I}$ is created by the dynamics of the demon, which also has an associated dissipation $\dot \Sigma_D$. Thus, one can define the thermodynamic efficiency for the creation of correlations as
\begin{equation}
    \eta_D = \frac{k_b \mathcal{\dot I}}{\dot \Sigma_D}.
\end{equation}
According to the second inequality in Eqs. \eqref{eq:local_second_laws_ss}, $\eta_D \leq 1$. Also, according 
to the first inequality in Eqs. \eqref{eq:local_second_laws_ss}, 
a positive information flow might be employed to compensate for a locally negative entropy production $\dot \Sigma_S \leq 0$. In those situations, the following thermodynamic efficiency can be assigned to that process
\begin{equation}
\eta_S = -\frac{\dot \Sigma_S}{k_b \mathcal{\dot I}},
\end{equation}
and we also have $\eta_S \leq 1$. Note that the net thermodynamic efficiency defined in Eq. \eqref{eq:global_eff} is just $\eta = \eta_S \eta_D$. 

\subsection{Scaling laws of information flows and efficiencies}

\begin{figure}
    \centering
    \includegraphics[scale=.6]{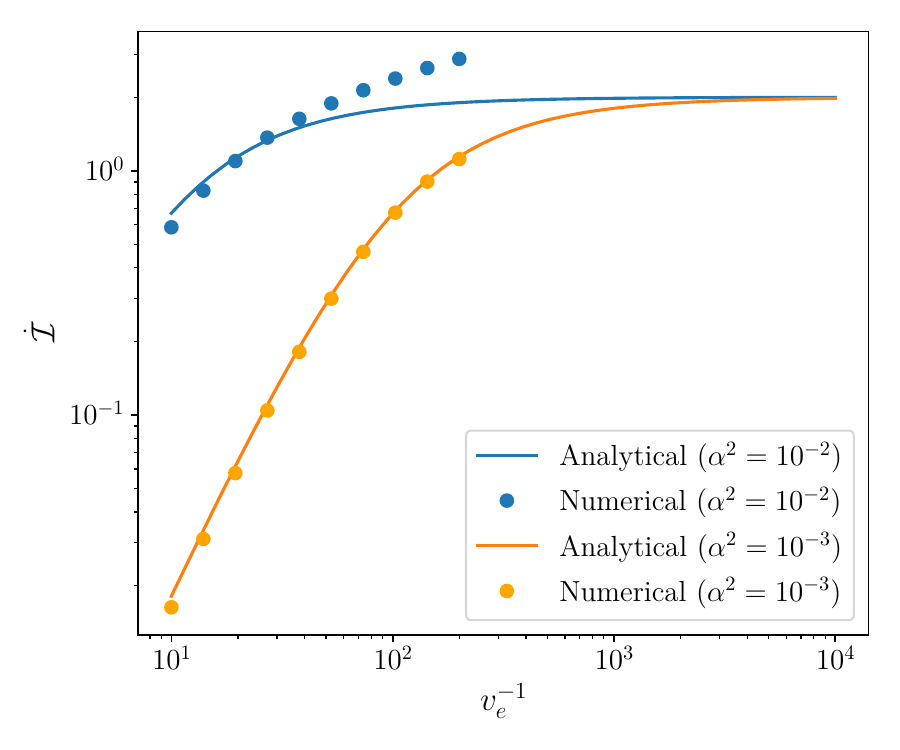}
    \caption{Information flow as a function of the scale parameter $\Omega = v_e^{-1}$ for the 
    scaling strategy $\Delta V_S =c v_e$ and $\Delta V_D = 2\log(1+2\alpha^2/cv_e)$, for different values of $\alpha^2$. The solid lines show the analytical result of 
    Eq. \eqref{eq:inf_flow}, while the points show numerical results obtained from the steady-state distribution and Eq. \eqref{eq:inf_flow_exact}.} 
    \label{fig:inf_flow}
\end{figure}

We now discuss how the different flows and efficiencies scale with respect to the scale parameter $\Omega = v_e^{-1}$. All other parameters fixed, we know that entropy flows $\dot \Sigma_{S/D}$ are extensive quantities (since the rates $\lambda_\pm^\text{n/p}(\bm{v})$ are extensive, and therefore the electric currents are). Also, we have seen that the information flow $\mathcal{\dot I}$ is intensive. This means that the efficiency $\eta_D$ scales as $1/\Omega$. It also means that $\dot \Sigma_S$ must become positive above some value of $\Omega$, since otherwise we would have $\eta_S > 1$, violating the first inequality in Eq. \eqref{eq:local_second_laws_ss}. As we have shown, this is indeed what happens in our electronic demon: rectification is impossible above a certain scale (see Eq. \eqref{eq:max_scale}).

It was shown in \cite{Freitas2022demon} that a way around the previous limitation is to make the voltage biases $\Delta V_S$ and $\Delta V_D$ to be scale dependent. In particular, if we take $\Delta V_S =c v_e \propto 1/\Omega$, and $\Delta V_D = 2\log(1+2\alpha^2/cv_e) \propto \log(\Omega)$, we obtain that $\dot \Sigma_S$ is scale independent and that $\dot \Sigma_D \propto \Omega^2\log(\Omega)$. Thus, the demon continues to work for any scale parameter, but with a net efficiency that decreases as $\eta = 1/\Omega^2 \log(\Omega)$. 
It is interesting to note, from Eq. \eqref{eq:inf_flow}, that even under this scaling strategy the information flow remains intensive, reaching the value $\mathcal{\dot I} = 2v_e/\tau_0$ for $v_e\to 0$ (recall that $\tau_0 \propto v_e$). Then the efficiency $\eta_S$ is also intensive an attains the limiting value:
\begin{equation}
    \lim_{v_e\to 0} \eta_S  = \frac{\alpha^2}{2},
\end{equation}
as can be seen from Eqs. \eqref{eq:inf_flow} and \eqref{eq:current_auto_linear}. In contrast,
the efficiency $\eta_D$ continuously decreases as  $\eta_D \propto 1/\Omega^2 \log(\Omega)$.
The behaviour of the information flow $\mathcal{\dot I}$ under the scaling strategy above is 
shown in Figure \ref{fig:inf_flow}, where the analytical result in Eq. \eqref{eq:inf_flow} is 
compared to numerical results obtained by computing the steady state distribution as in 
Section \ref{sec:auto_stoch_dyn} and using Eq. \eqref{eq:inf_flow_exact} for the information 
flows. The numerical results are limited to low values of the scale parameter, since otherwise 
the exact computation of the steady state distribution becomes too expensive. We see that the 
analytical result of Eq. \eqref{eq:inf_flow} is only accurate for low values of $\alpha^2$. 
The reason is that the information flows are highly sensitive to the non-Gaussian nature of the exact steady-state distribution.

In relation to this last observation, it is important to discuss the role that the LD principle plays in our calculation, and how is it affected by the previous scaling strategy. The LD principle and the associated rate function was only employed here as a computational tool to extract the Gaussian covariance matrix. It provides the correct Gaussian moments (related to the curvature of the rate function around its minimum) for arbitrary parameters $\Delta V_S$ and $\Delta V_D$ (as long as $\alpha^2<1$ and the scale parameter is large enough). However, one should note that the LD principle ceases to be valid for the scaling strategy $\Delta V_S \propto 1/\Omega $ and $\Delta V_D = \log(\Omega)$, since the transition rates are no longer extensive. One way to see this is to note that the variance of the voltages does not scale as $1/\Omega$, which is the central limit theorem scaling expected from the the LD principle. A related consequence is that a Gaussian approximation (that is, the truncation of $-\log(P_\text{ss}(\bm{v}))$ to second order around its minimum) is not justified anymore in the limit $\Omega \to \infty$, except if other conditions ($\alpha^2 \ll 1$ in our case) are met. This explains the discrepancies in Figure \ref{fig:inf_flow}.

Finally, we note that we have only considered the situation in which system and demon have the same scale. This is natural in this setting, since the output capacitance of the system(demon) inverter is dominated by the gate-body capacitance of the transistors in the demon(system) inverter. A similar analysis holds for the case in which only the demon inverter is scaled up. If the system is scaled up, while the demon scale is held constant, then the demon power still needs to be scaled up in order for it to be able to drive the input capacitance of the system inverter. In other words, when the demon is small compared to the system, the feedback process becomes the limiting factor in our circuit.

\section{Conclusion}
We have presented a in-depth study of the electronic implementation of an autonomous Maxwell's demon proposed in \cite{Freitas2022demon}. We have adapted the thermodynamic theory of information flows developed in \cite{horowitz2014} to systems displaying a macroscopic limit and applied it to the electronic demon. This allowed us to define detailed thermodynamic efficiencies for the processes creating and consuming correlations between system and demon, in terms of an information flow that was explicitly computed. Also, we have shown that the information flow is an intensive quantity in any autonomous bipartite system with a macroscopic limit, which implies that any such system will stop working as a demon above a finite scale. Implications in chemical and biological systems \cite{Ehrich2022} will be explored elsewhere.

\section{Acknowledgments}
This research was supported by the project INTER/FNRS/20/15074473 funded by F.R.S.-FNRS (Belgium) and FNR (Luxembourg),
and by the FQXi foundation project FQXi-IAF19-05.

\bibliographystyle{unsrt}
\bibliography{references.bib}

\end{document}